\documentclass[a4paper,12pt]{article}
\pdfoutput=1 
\usepackage{jcappub-mod}
\usepackage[T1]{fontenc}
\usepackage[normalem]{ulem}

\title{\textbf{Testing Non-minimally Coupled BEC Dark Matter with Gravitational Waves}}

\author[a,b]{Dimitar Ivanov,}

\author[a,b,c]{Stefano Liberati}

\affiliation[a]{SISSA - International School for Advanced Studies, 
via Bonomea 265, 34136 Trieste, Italy.}

\affiliation[b]{IFPU, Via Beirut 2, 34014, Trieste, Italy.}

\affiliation[c]{INFN, sezione di Trieste, via Valerio 2, Trieste, Italy.}

\emailAdd{divanov@sissa.it}
\emailAdd{liberati@sissa.it} 

\abstract{We study the phenomenology associated to non-minimally coupled dark matter. In particular, we consider the model where the non-minimal coupling arises from the formation of relativistic Bose-Einstein condensates in high density regions of dark matter \cite{bec}. This non-minimal coupling is of Horndeski type and leads to a local modification of the speed of gravity with respect to the speed of light. Therefore we can constrain the model by using the joint detection of $\mathrm{GW170817}$ and $\mathrm{GRB170817A}$. We show that the constraints obtained in this way are quite tight, if the dark matter field oscillates freely, whereas they are substantially weakened, if the oscillations are damped by the non-minimal coupling.}


\begin{document}

\maketitle

\flushbottom


\section{Introduction}
\label{introduction}

The proposal that Dark Matter (DM) can form large Bose-Einstein Condensates (BEC) has been discussed in several papers \cite{harko, cosmic, cusp, growth, shapiro}.  While all of them deal with a non-relativistic BEC in a Newtonian potential, it was studied in \cite{bec} the case of a relativistic BEC in a curved spacetime. It was there conjectured that as the condensation occurs and a characteristic length scale - the healing length of the condensate, develops, a non-minimal coupling between the Bose field and the spacetime metric forms. The reasoning is that since the healing length can be macroscopic and comparable to the length scale set by the curvature of spacetime, this implies that the condensate can probe the geometry non-locally and the minimal coupling principle does not hold any more. The formation of the non-minimal coupling modifies the behaviour of DM on small (galactic scales) allowing to improve some of the small-scale problems faced by the Cold Dark Matter (CDM) paradigm such as the Core-Cusp Problem. At the same time it retrieves all the successes of CDM on large scales.

The basic idea behind the model is that at initial stages of the evolution of the Universe, the temperature of the Dark Matter, which can be defined for a weakly self-coupled field, is much higher than the critical temperature below which the condensate forms. As the Universe expands, the temperature of the field drops and at the same time different regions start collapsing under their own gravitational weight, as structure formation begins. As these local gravitational potential wells form and gradually become deeper with time, the critical temperature within them rises and at some point becomes equal to the decreasing temperature of the Bose field. At this point condensation can occur within the potential wells and a non-minimal coupling forms, which locally modifies the small-scale behaviour of the Bose field. At the same time on larger scales, outside the potential wells the Bose field is minimally coupled and behaves just like CDM. Thus the advantage of this model is that it preserves the CDM-like behaviour on large scales while predicting deviations from CDM on small scales consistently with observations. 

An interesting and very important question is how the model can be tested. One consequence of the non-minimal coupling is that it causes gravitational waves to propagate with a speed different from that of photons. As a consequence, two signals - gravitational and electromagnetic, emitted simultaneously would accumulate a difference in their arrival times as seen by an observer far away. The recently detected Gravitational Wave (GW) event $\mathrm{GW170817}$ followed by the Gamma Ray Burst (GRB) event $\mathrm{GRB170817A}$ approximately 1.7s later is thought to have originated during a Neutron Star - Neutron Star (NS-NS) merger event located in the outskirts of an elliptical galaxy approximately $40 \mathrm{Mpc}$ away from us \cite{event}. The short arrival time difference between the two types of waves has been used to eliminate and constrain various models of Dark Energy \cite{ferreira, creminelli, zuma}. That is because the majority of possible non-minimal couplings between the gravitational metric and a scalar field would predict anomalous gravitational wave propagation and therefore - a sizeable difference in the arrival times. 

The observation has also been used to exclude some dark matter emulator models i.e. models which modify gravity so as to eliminate the need for Dark Matter and to put sharp constraints on possible violations of the Weak Equivalence Principle \cite{boran, wei, wang}. However, the implications of this observation for theories of Dark Matter where the latter genuinely exists but interacts non-trivially with the gravitational field, have, to the best of our knowledge, never been discussed.

In this paper we investigate whether the GW - GRB event is consistent with the model of non-minimally coupled BEC as DM. Since the non-minimal coupling is active only in regions with strong gravitational fields such as galaxy and cluster halos, significant time lag of electromagnetic waves with respect to gravitational waves would accumulate only as the waves pass through these regions. We estimate the predicted total arrival time difference and compare it to the observed one. Our aim is to check whether the model is compatible or not with the observation and to constrain the parameter space of the model.

There are two main free parameters that one needs to constrain - $m$ - the mass of the scalar field which condenses and forms the Dark Matter, and $\varsigma$ - the coupling constant of the non-minimal coupling between the Dark Matter and the metric. However, since the model developed in~\cite{bec} is a phenomenological one, there are several uncertainties that we also need to parametrize. 

More precisely, since we do not have a complete mathematical theory of condensation in a curved spacetime, we do not know how strong the gravitational field needs to be in order for the condensation to occur. As a result, we do not know whether dark matter is in a condensate phase only inside galaxy halos or also inside cluster halos. We will parametrize this uncertainty by introducing a free parameter $\beta$. 

Furthermore, we know that when the field is not in the condensate phase it has to oscillate in order to behave as Cold Dark Matter. However, in the condensate phase the non-minimal coupling will change the evolution of the scalar field and, as we are going to argue later, it might actually damp the oscillations. In order to parametrize our uncertainty of whether the field oscillates or not, we need to introduce two other free parameters - $\gamma_{1}$ and $\gamma_{2}$.  We will compute the total predicted arrival time difference in terms of these five free parameters and then will investigate how it depends on them and consider some limiting cases.

Throughout the paper we work at an order of magnitude level and we make several simplifying approximations and assumptions that will be clearly spelled out later. Any analysis beyond that would involve the running of numerical simulations which goes beyond the scope of the present work.

The structure of the paper is the following. In Section \ref{model} we present a brief overview of the model and we consider under what conditions it could solve the Core-Cusp problem. In Section \ref{speed} we look at how the non-minimal coupling between DM and the metric modifies the speed of GWs considering both the cases of a timelike and a spacelike gradient of the DM field. In Section \ref{time} we compute the total predicted arrival time difference between GWs and GRBs in terms of the five free parameters of the model. Finally, we draw our conclusions in Section \ref{conclusion}.

Throughout the paper we use a convention in which $c = \hbar = 1$ unless stated otherwise and take the signature of the metric to be $(-1, 1, 1, 1)$.


\section{Relativistic BEC as DM}
\label{model}


\subsection{Overview of the Model}

A relativistic Bose-Einstein Condensate (BEC) is a relativistic gas of bosons in which most of the particles have undergone a phase transition and are in a condensate phase characterised by a vanishing wave number: $k=0$. The starting relativistic action in flat spacetime is
\begin{equation}
S = - \int \left[ \frac{1}{2} \partial_{\mu} \hat{\phi}^{\dagger} \partial^{\mu} \hat{\phi} + \frac{1}{2} m^2 \hat{\phi}^{\dagger} \hat{\phi} + U(\hat{n}, \lambda) \right] d^4 x, \label{rbec}
\end{equation}
where $\hat{\phi}(\vec{x}, t)$ is a relativistic scalar Bose field operator, $\hat{n}$ is defined by $\hat{n} = \hat{\phi}^{\dagger} \hat{\phi}$ (in the non-relativistic theory this has the interpretation of a number density, while in the relativistic theory it is simply a quantity related to the mass density $\hat{\rho}$ by $\hat{\rho} = m^2 \hat{n}$), $\lambda$ is a dimensionless coupling constant and $U(\hat{n}, \lambda)$ is a self-interaction term of the form
\begin{equation}
U(\hat{n}, \lambda) = \frac{\lambda}{2} \hat{n}^2.
\end{equation}
One could also add an external potential $V(\vec{x}, t)$ to the action \eqref{rbec}. The condensation occurs below a critical temperature $T_c$. In the condensate phase the Bose field can be split into a classical complex scalar field $\phi$ describing the ground state, and quantum excitations $\hat{\varphi}$:
\begin{equation}
\hat{\phi} = \phi (\hat{1} + \hat{\varphi}).
\end{equation}
The field $\phi$ obeys a relativistic Klein-Gordon equation of the form
\begin{equation}
\square \phi - m^2 \phi - U^{\prime} \phi = 0,
\end{equation}
where the prime denotes derivative with respect to the number density defined above. In the non-relativistic limit this reduces to a non-linear Schroedinger-like equation called the Gross-Pitaevski equation and for this reason $\phi$ is often called the ``wave function'' of the condensate. It is useful to express the condensate wave function and its complex conjugate in terms of the hydrodynamic variables $\rho = m^2 \phi^{*} \phi$ and $u^{\mu}$ in the so-called Madelung representation:
\begin{equation}
\phi \equiv \frac{1}{m} \sqrt{\rho} e^{i \theta},
\end{equation}
\begin{equation}
u^{\mu} \equiv \frac{1}{m} \nabla^{\mu} \theta. \label{velocity}
\end{equation}
The 4-velocity $u^{\mu}$ is in general not normalised. One can instead define the normalised velocity $v^{\mu} \equiv u^{\mu} / \sqrt{- u^{\mu} u_{\mu}}$ but for our purposes this is not necessary and we will keep using $u^{\mu}$ in the subsequent formulae. \footnote{The quantity $\rho$ is obtained by taking the vacuum expectation value of the operator $\hat{\rho} = m^2 \hat{n}$ and can be interpreted as the rest mass density in the BEC frame. One can check that this is the correct interpretation by deriving the stress-energy tensor of the scalar field and contracting it with $v^{\mu}$ in order to obtain the energy density.} With these definitions the Klein-Gordon equation can be shown to be equivalent to two equations in terms of $\rho$ and $u^{\mu}$:
\begin{equation}
\nabla_{\mu} (\rho u^{\mu}) = 0, \label{cont}
\end{equation}
\begin{equation}
u^{\mu} u_{\mu} = -1 + \Big( - 2 U^{'}(\rho) + \frac{1}{m^2} \frac{\square \sqrt{\rho}}{\sqrt{\rho}} \Big) \label{frank}
\end{equation}
where $U^{\prime}(\rho)$ is the derivative of the self-interaction potential $U$ with respect to the mass density $\rho$ and $V_{q} \equiv \frac{1}{m^2} \frac{\square \sqrt{\rho}}{\sqrt{\rho}}$ is the quantum potential. The first equation is the continuity equation while the second is an equation for the norm of $u^{\mu}$. It shows that the 4-velocity $u^{\mu}$ is generally not normalised but that for CDM for which both $U(\rho) =0$ and $V_{q} = 0$ it reduces to the normalised velocity. Eqn. \eqref{frank} can be turned into a dynamical equation by taking the covariant derivative:
\begin{equation}
u^{\mu} \nabla_{\mu} u_{\nu} = - \partial_{\nu} \Bigg[ U^{\prime}(\rho) \Big( 1 - \xi^2 \frac{\square \sqrt{\rho}}{\sqrt{\rho}} \Big) \Bigg], \label{euler}
\end{equation}
where $\xi^2 \equiv 1/(2m^2 U^{\prime}(\rho))$ is the healing length which is the length scale over which the density of the condensate returns to its bulk value when perturbed locally. Notice the appearance of a third derivative on the RHS of the equation due to the extra covariant derivative that we took.

If we consider a uniform square potential well, the density must vanish at the boundary and become constant towards the centre of the square well. In that case, the healing length would correspond to the length scale over which the density rises from $0$ to that constant value (see \cite{pethick} for more details). If the typical length and time scales of variation of the density $\rho$ are much larger than $\xi$, then we can ignore the quantum pressure term in \eqref{euler}, in which case \eqref{cont} and \eqref{euler} are completely equivalent to the continuity and Euler equations of a relativistic perfect fluid.

So far we have considered a relativistic BEC in a flat spacetime. What happens, if instead the condensation occurs in a curved spacetime? We saw that the process of condensation is characterised by the formation of a typical length scale - the healing length $\xi$. If this length scale is comparable to the length scale set by the curvature of spacetime, i.e. the length scale over which the geometry of spacetime starts deviating from Minkowski, then the condensate would become sensitive to the global geometry. Thus one would expect the formation of a non-minimal coupling - a direct coupling between the curvature and the Bose field with a coupling constant directly related to the healing length of the BEC. Unfortunately, in the absence of a rigorous mathematical theory of condensation in a curved spacetime, one cannot say for sure what the form of this coupling will be. However, we can impose several requirements that narrow the range of possible couplings.

The requirement to have second order field equations leads to a coupling which is a subclass of the Horndeski Lagrangian \cite{modgrav}. We also require that the coupling contains a dimensionful coupling constant (thus $\phi R$ would not work for example) in order to justify why this coupling is present for DM but not for baryons (because baryons have no macroscopic coherence length scale, while DM has - the healing length). Furthermore, in \cite{mond, extended} it was shown that a coupling of the form $G_{\mu \nu} \nabla^{\mu} \phi \nabla^{\nu} \phi$ in the Jordan frame leads to an effective modified gravity behaviour on small scales which, in the Einstein frame, causes baryons to propagate on an effective metric different from that on which DM propagates and thus leads to an effective MOND-like phenomenology. Based on this reasoning it was conjectured in \cite{bec} that the non-minimal coupling developed during the process of condensation is $L^2 G_{\mu \nu} \nabla^{\mu} \phi^{\dagger} \nabla^{\nu} \phi$ where L is a coupling constant with dimensions of length. This coupling satisfies all the above requirements and in Appendix \ref{equivalence} we show that it is equivalent to a subpart of the Horndeski Lagrangian.

Thus the action of the Bose field in the condensate phase is taken to be a sum of three terms
\begin{equation}
S = S_{EH} + S_{\phi} + S_{NMC} , \label{action}
\end{equation}
where
\begin{equation}
S_{EH} = \frac{1}{16 \pi G} \int R \sqrt{-g} d^4 x
\end{equation}
is the standard Einstein-Hilbert term,
\begin{equation}
S_{\phi} = - \int \left[ \frac{1}{2} g^{\mu \nu} \nabla_{\mu} \hat{\phi}^{\dagger} \nabla_{\nu} \hat{\phi} + \frac{1}{2} m^2 \hat{\phi}^{\dagger} \hat{\phi} + U(\hat{n}, \lambda_i) \right] \sqrt{-g} d^4 x
\end{equation}
is the minimally coupled part of the Bose field and
\begin{equation}
S_{NMC} = \int L^2 G_{\mu \nu} \nabla^{\mu} \hat{\phi}^{\dagger} \nabla^{\nu} \hat{\phi} \sqrt{-g} d^4 x
\end{equation}
is the non-minimally coupled part of the Bose field. Varying the above action with respect to $g^{\mu \nu}$ gives the gravitational field equation
\begin{equation}
G_{\mu \nu} = 8 \pi G \big[ T^{\phi}_{\mu \nu} + T^{NMC}_{\mu \nu} \big], \label{grav}
\end{equation}
where  the explicit form of $T^{\phi}_{\mu \nu}$ and $T^{NMC}_{\mu \nu}$ is provided in \cite{bec}.  \footnote{We must stress that the relation between the non-condensate phase where the field is minimally coupled and the condensate phase where a non-minimal coupling exists is not equivalent to the relation between a UV theory and an IR theory in the EFT sense. In particular, both regimes exist below the Planck scale and are not related to each other by a RG flow, rather by a thermodynamic phase transition.} It is useful to manipulate this equation by imposing both a fluid limit and a Newtonian limit. 

As we saw before, for the fluid limit we need to express $T^{\phi}_{\mu \nu}$ and $T^{NMC}_{\mu \nu}$ in terms of the fluid variables $\rho$ and $u^{\mu}$ and we assume that $\rho$ varies slowly enough for ignoring the quantum pressure term in \eqref{euler}. 

For the Newtonian limit we expand the metric as the flat spacetime metric plus a first-order scalar mode perturbation $\Phi$ and we assume small time gradients and small velocities compared to the speed of light \cite{bec}. At the end the field equation \eqref{grav} reduces to a modified Poisson equation of the form
\begin{equation}
\nabla^2 \Phi = 4 \pi G (\rho - L^2 \nabla^2 \rho). \label{poisson}
\end{equation}
 After this succinct recap of the model let us now discuss under what conditions it could be able to solve one of the long standing problems of CDM models, namely the Core-Cusp problem.


\subsection{The Core-Cusp Problem}

Numerical simulations with CDM predict that the density profile of a virialised Dark Matter halo is given by the Navarro-Frenk-White (NFW) profile \cite{salucci, locas}
\begin{equation}
\rho_{NFW}(r) = \frac{\rho_{s}}{\frac{r}{r_s} (1 + \frac{r}{r_s})^2}, \label{nfw}
\end{equation}
where $r$ is the radius from the centre of the halo and $\rho_{s}$ and $r_{s}$ are constants specific for the galaxy of interest (thus they are free parameters in the simulation). The NFW profile has a cusp at the centre. However, observations favour density profiles with a core at the centre. For example, the observed circular velocities in spirals, dwarf disks and low
surface brightness systems are well fitted by an empirical Burkert profile \cite{salucci, burkert}
\begin{equation}
\rho_{B}(r) = \frac{\rho_0}{(1 + \frac{r}{r_0})(1 + (\frac{r}{r_0})^2)}, \label{Burkert}
\end{equation}
 where $r_0$ is the radius of the core and $\rho_0$ is the density at the centre. This discrepancy between CDM simulations, which predict a cusp, and observations, which favour a core, is called the Core-Cusp problem.

In the numerical simulations which lead to the cuspy profile \eqref{nfw} Dark Matter is modelled as dust i.e. a pressureless collection of particles interacting only via standard (Newtonian) gravity. However, in our model Dark Matter consists of a Bose-Einstein Condensate which gravitates in the non-relativistic limit via the modified Poisson equation \eqref{poisson}. This would lead to a different prediction for the equilibrium density profile. In order to find it, we need to solve a set of three coupled partial differential equations in terms of three variables $\Phi$, $\rho$ and $\vec{v}$. The first is, of course, the modified Poisson equation \eqref{poisson}, the other two are the fluid equations \eqref{cont} and \eqref{euler} in the non-relativistic limit:
\begin{equation}
\nabla^2 \Phi = 4 \pi G (\rho - L^2 \nabla^2 \rho), \label{poi}
\end{equation}
\begin{equation}
\frac{\partial \rho}{\partial t} + \vec{\nabla} . (\rho \vec{v}), = 0
\end{equation}
\begin{equation}
m \frac{\partial \vec{v}}{\partial t} = -m (\vec{v} . \vec{\nabla}) \vec{v} - m\vec{\nabla} \Phi + \vec{\nabla} \Big[ U^{'} (\rho) \Big( 1 - \xi^2 \frac{\nabla^2 \sqrt{\rho}}{\sqrt{\rho}} \Big) \Big]. \label{vel}
\end{equation}
However, solving these three equations analytically is not possible. A numerical simulation would be required in order to evolve $\rho$ and $\vec{v}$ from some specified initial conditions. However, we can still argue from the structure of these equations that they would tend to smooth out any cusps. In particular, the quantum pressure term $\xi^2 \vec{\nabla}   \Big( U^{'} (\rho)  \frac{\nabla^2 \sqrt{\rho}}{\sqrt{\rho}} \Big)$ in \eqref{vel} would prevent the building up of large gradients of the density. 

We can actually find an approximate relation between $\xi$ and $m$ such that the cusp within galaxy halos is smoothed out. For that we consider the regime after the halo has relaxed to its equilibrium density profile. In that regime the gradients of the density and the velocities are negligible. This means that in \eqref{poi} and \eqref{vel} we can neglect both the term from the non-minimal coupling proportional to $L^2$ and the quantum pressure term proportional to $\xi^2$. It is shown in \cite{harko} that solving the fluid equations in this regime and looking for static solutions $\vec{v} = 0$ leads to the following density profile:
\begin{equation}
\rho(r) = \rho_{0} \frac{\mathrm{sin} (kr)}{kr}, \label{sin}
\end{equation}
where $k = \sqrt{\frac{G m^3}{\hbar^2 a}}$ and $a$ is the scattering length. The first zero of \eqref{sin} i.e. the smallest $r$ for which $\rho$ vanishes, occurs at
\begin{equation}
R = \pi \sqrt{\frac{\hbar^2 a}{G m^3}}, \label{rad}
\end{equation}
which we take to be the same as the radius of the halo $R = R_{H}$. On the other hand, for a non-relativistic BEC the healing length is related to the scattering length by \cite{analog}
\begin{equation}
\xi = \frac{1}{\sqrt{4 \pi a n}}, \label{heal}
\end{equation}
where $n \approx \bar{\rho}/m$ is the number density of particles in the condensate phase. Combining \eqref{rad} and \eqref{heal} allows us to express the healing length as a function of the mass:
\begin{equation}
\xi = \frac{\hbar}{m} \frac{\pi}{R} \frac{1}{\sqrt{4 \pi G \bar{\rho}}}. \label{healcore}
\end{equation}
This relation is essentially obtained by requiring that the density distribution of the BEC in the non-relativistic minimally coupled limit is of the same size as the size of the typical galaxy halo. Since the density distribution of the BEC does not have the cusp at the centre, one can think of this as the healing length necessary to solve the Core-Cusp problem (of course, as mentioned before in order to really demonstrate that the Core-Cusp problem is solved, one would have to run numerical simulations). 

In order to find the coupling constant $L$ which favours a smooth core, we need to find a relation between $L$ and $\xi$. Since the non-minimal coupling is present when $\xi$ is of the same order as $l_{c}$ the length scale of the curvature, it is the ratio $\xi/l_{c}$ which controls the strength of the non-minimal coupling. Since, $l_{c}$ already appears in the Einstein tensor, $L$ has to be of the same order as $\xi$. Indeed, taking $L = \xi$ we have that 
\begin{equation}
\xi^2 G_{\mu \nu} \nabla^{\mu} \phi \nabla^{\nu} \phi \sim \Big( \frac{\xi}{l_{c}} \Big) g_{\mu \nu} \nabla^{\mu} \phi \nabla^{\nu} \phi,
\end{equation}
and therefore the term becomes important only when $\xi \sim l_{c}$. Therefore, from now on we will fix $L = \xi$. We need to stress that this identification between the two length scales does not follow from the mathematics of the model or from any rigorous mathematical theory of condensation in a curved spacetime. Instead, it is based on intuition and the desire to consider the simplest possible model. So even though we assume that for now, one must be ready to entail the possibility that $L$ and $\xi$ could be related in some more complicated way. Writing the coupling constant as $\varsigma \equiv 1/L$ gives a relation between $\varsigma$ and $m$:
\begin{equation}
\varsigma = m \frac{R}{\pi} \sqrt{4 \pi G \rho}. \label{mu}
\end{equation}
Equation \eqref{mu} gives the coupling constant $\varsigma$ which favours a smooth core as a function of the mass $m$.

\subsection{Field oscillations and CDM limit}

The complex scalar field $\phi$ should behave as CDM, i.e. pressureless dust, on large scales in order to fit with cosmological observations. As we noted in Section \ref{model}, this implies that in this case the 4-velocity $u^{\mu}$ is normalized and therefore there exists a frame where it takes the form  $u^{\mu} = (1, 0, 0, 0)$. Because of the isotropy on large scales, the time direction in $u^{\mu}$ necessarily coincides with the cosmic time direction as defined in FLRW. From \eqref{velocity} this implies that $\dot{\theta} = m$ and therefore the scalar field can be written as
\begin{equation}
\phi = A(r) \mathrm{e}^{i (mt + \alpha_{0})} \label{radially}
\end{equation}
where $A \equiv | \phi |$ is the amplitude of oscillations - we write it as $A(r)$ to emphasise that it depends on $r$ only. The energy of oscillations is
\begin{equation}
\rho = \frac{1}{2} | \dot{\phi} |^2 + \frac{1}{2} m^2 | \phi |^2 = m^2 A^2 = m^2 \phi^{*} \phi
\end{equation}
consistent with the hydrodynamic definition of the density.

For later convenience, it would also be useful to split the complex field in terms of its real and imaginary components:
\begin{equation}
\phi = \phi_{1} + i \phi_{2}
\end{equation}
where each component oscillates according to:
\begin{equation}
\phi_{1} = A(r) \mathrm{cos} (mt + \alpha_{0})
\end{equation}
\begin{equation}
\phi_{1} = A(r) \mathrm{sin} (mt + \alpha_{0})
\end{equation}

These oscillations guarantee that the pressure averages out to zero and therefore that the field behaves as CDM on large scales. We will treat the CDM-like behaviour as a zero-level approximation for the evolution of the field inside galaxies and clusters. Later we will parametrise deviations from this behaviour by taking into account the possible damping due to the non-minimal coupling. Having reviewed the model of Dark Matter as a non-minimally coupled BEC, we next turn to the question of how gravitational wave propagation is modified within this model.


\section{Gravitational Waves as a Probe for a Non-minimal Coupling between Curvature and Matter}
\label{speed}

It is well known that in several modified theories of gravity, in particular Horndeski Scalar-Tensor theories, gravitational waves propagate with a speed different from that of light. It is therefore of no surprise that in our model, where a non-minimal coupling forms after a phase transition, this phenomenon is also present. We now outline the mathematical procedure to calculate the speed of gravitational waves. We choose to work at the level of the field equations, although another possibility would have been to work at the level of the action (see, for example,\cite{tens} and  \cite{bettoni} for such an approach).

Starting from the gravitational part of the action \eqref{action} expanded in terms of the condensate wave function $\phi$: 
\begin{equation}
S_{EH} =  \int \Big[ \frac{1}{16 \pi G} R + L^2 G_{\mu \nu} \nabla^{\mu} \phi^{*} \nabla^{\nu} \phi \Big] \sqrt{-g}  d^4 x
\end{equation}
it takes a tedious but straightforward calculation to obtain the gravitational field equation
\begin{align}
\frac{1}{16 \pi G} G_{\mu \nu} &+ L^2 \Big[ -\frac{1}{2} X G_{\mu \nu} + \frac{1}{4} \square \phi^{*} \square \phi g_{\mu \nu} - \frac{1}{4} \nabla_{\alpha} \nabla_{\beta} \phi^{*}  \nabla^{\alpha} \nabla^{\beta} \phi g_{\mu \nu} \nonumber \\
& - \frac{1}{4} R \nabla_{\mu} \phi^{*} \nabla_{\nu} \phi  + \frac{3}{2} \nabla^{\beta} \nabla_{\mu} \phi^{*} \nabla_{\beta} \nabla_{\nu} \phi - \frac{1}{2} \nabla_{\mu} \nabla_{\nu} \phi^{*} \square \phi + \frac{1}{2} \nabla^{\lambda} \phi^{*} \nabla_{\mu} \phi R_{\lambda \nu} \nonumber \\
& + \frac{1}{2} \nabla^{\lambda} \phi^{*} \nabla_{\nu} \phi R_{\lambda \mu} + \frac{1}{2} R_{\mu \beta \nu \lambda} \nabla^{\beta} \phi^{*} \nabla^{\lambda} \phi - \frac{1}{2} R_{\lambda \beta} \nabla^{\lambda} \phi^{*} \nabla^{\beta} \phi g_{\mu \nu} + c.c. \Big] = 0.
\end{align}
where c.c. stands for the complex conjugate. It reduces to the Einstein Field equation when the non-minimal coupling constant is sent to zero: $L \rightarrow 0$, as expected. We expand both the metric and the scalar field (condensate wave function) in terms of a background plus a first order perturbation, use local flatness to write the metric background as Minkowski
\begin{equation}
g_{\mu \nu} = \eta_{\mu \nu} + h_{\mu \nu},
\end{equation}
\begin{equation}
\phi = \bar{\phi} + \varphi,
\end{equation}
impose the de Donder gauge
\begin{equation}
\partial_{\mu} h^{\mu \nu} = \frac{1}{2} \partial_{\nu} h,
\end{equation}
and require that the trace vanishes
\begin{equation}
h \equiv \eta^{\mu \nu} h_{\mu \nu} = 0.
\end{equation}
The latter is a restriction of the solution, not a gauge fixing. Since we are interested only in the propagation of tensor modes, and since tensor and scalar modes decouple from each other in linear theory, we can without loss of generality impose the above restriction, thereby removing any scalar modes involving the trace from the equation \cite{waves}. The final result is the linearised field equation\footnote{Some steps of that and the previous computation were performed using the xAct package of Mathematica.}
\begin{align}
& \frac{1}{16 \pi G} \square h_{\mu \nu} - L^2 \Big[ \frac{1}{4} \square h_{\mu \nu} \partial_{\rho} \bar{\phi}^{*} \partial^{\rho} \bar{\phi} - \frac{1}{2} \square h_{\rho \mu} \partial_{\nu} \bar{\phi}^{*} \partial^{\rho} \bar{\phi} \nonumber \\
&- \frac{1}{2} \square h_{\rho \nu} \partial_{\mu} \bar{\phi}^{*} \partial^{\rho} \bar{\phi} + \frac{1}{2} \square h_{\rho \sigma} \partial^{\rho} \bar{\phi}^{*} \partial^{\sigma} \bar{\phi} \eta_{\mu \nu} + \frac{1}{2} ( - \partial_{\mu} \partial_{\nu} h_{\rho \sigma} \nonumber \\
&+ \partial_{\mu} \partial_{\rho} h_{\nu \sigma} + \partial_{\nu} \partial_{\rho} h_{\mu \sigma} - \partial_{\rho} \partial_{\sigma} h_{\mu \nu} ) \partial^{\rho} \bar{\phi}^{*} \partial^{\sigma} \bar{\phi} + c.c. \Big] = 0. \label{linear}
\end{align}
The real fields $\phi_{1}$ and $\phi_{2}$ appear on the same footing in the Lagrangian and thus satisfy similar equations of motion. Therefore, the gradients of both $\phi_{1}$ and $\phi_{2}$ must be of the same nature - spacelike or timelike. Thus from now on we will only speak about the gradient of $\bar{\phi}$. There are two cases that we need to consider - first, when the gradient of $\bar{\phi}$ is timelike, and second when it is spacelike.

\subsection{The Case of a Timelike Gradient of $\bar{\phi}$}

If the gradient of $\bar{\phi}$ is timelike, we can boost to a frame where it has the form \footnote{Technically, we need to do this for both components of the complex field. However, since the evolution of the two components is similar, the same boost would bring both components to approximately the same form.}
\begin{equation}
\bar{\phi}_{\mu} = (\dot{\bar{\phi}}, 0, 0, 0).
\end{equation}
We perform a $3+1$ decomposition of the perturbation $h_{\mu \nu}$ (again we are interested only in the propagation of tensor modes):
\begin{equation}
h_{00} = 0, \qquad h_{0i} = h_{i0} = 0, \qquad \partial^i h_{ij} = \delta^{ij} h_{ij} = 0.
\end{equation}
This allows to rewrite \eqref{linear} in the form of a wave equation
\begin{equation}
- \ddot{h}_{ij} + c_g^2 \nabla^2 h_{ij} = 0,
\end{equation}
from where it is straightforward to extract the speed of propagation of gravitational waves - $c_g$:
\begin{equation}
c_g^2 = \frac{1 + 8 \pi G L^2 | \dot{\bar{\phi}} |^2}{1 - 8 \pi G L^2 | \dot{\bar{\phi}} |^2}. \label{sp1}
\end{equation}

\subsection{The Case of a Spacelike Gradient of $\bar{\phi}$}

If the gradient of $\bar{\phi}$ is spacelike, then we can boost to a frame where it has the form
\begin{equation}
\bar{\phi}_{\mu} = (0, \vec{\nabla} \bar{\phi}).
\end{equation}
We separate the space gradient into components parallel and perpendicular to the propagation vector $\vec{k}$:
\begin{equation}
|\vec{\nabla} \bar{\phi}|^2 = |\bar{\phi}_{\parallel}|^2 + |\bar{\phi}_{\bot}|^2.
\end{equation}
Performing the $3+1$ decomposition as before and considering a single mode of wave vector $\vec{k}$ and frequency $\omega$,
\begin{equation}
h_{ij} = A_{ij} e^{i(\vec{k}.\vec{x} - \omega t)},
\end{equation}
allows to find the dispersion relation
\begin{equation}
\omega^2 - c_g^2 k^2 = 0,
\end{equation}
from where, again, it is straightforward to read the speed of propagation:
\begin{equation}
c_g^2 = \frac{ 1 + 8 \pi G L^2 | \bar{\phi}_{\parallel} |^2 + 8 \pi G L^2 | \bar{\phi}_{\bot} |^2}{ 1 - 8 \pi G L^2 | \bar{\phi}_{\parallel} |^2 + 8 \pi G L^2 | \bar{\phi}_{\bot} |^2}. \label{sp2}
\end{equation}
One can also rewrite the last formula in terms of the angles $\iota_{1}$ and $\iota_{2}$ between the wave vector $\vec{k}$ and the gradient of each component of the complex scalar field, $\vec{\nabla} \bar{\phi}_{1}$ and $\vec{\nabla} \bar{\phi}_{2}$:
\begin{equation}
c_g^2 = \frac{1+ 8 \pi G L^2 |\vec{\nabla} \bar{\phi}_{1}|^2 + 8 \pi G L^2 |\vec{\nabla} \bar{\phi}_{2}|^2}{1 - 8 \pi G L^2 |\vec{\nabla} \bar{\phi}_{1}|^2 \mathrm{cos}(2 \iota_{1}) - 8 \pi G L^2 |\vec{\nabla} \bar{\phi}_{2}|^2 \mathrm{cos}(2 \iota_{2})}. 
\end{equation}
In our case the scalar field is only radially dependent, Eqn. \eqref{radially}, which implies that the two gradients are aligned $\iota_{1} = \iota_{2} = \iota$, so the expression for the speed reduces to
\begin{equation}
c_g^2 = \frac{1+ 8 \pi G L^2 |\vec{\nabla} A|^2}{1 - 8 \pi G L^2 |\vec{\nabla} A|^2 \mathrm{cos}(2 \iota)}.  \label{sp21}
\end{equation}

A very important point to note is that Equations \eqref{sp2} and \eqref{sp21} imply that the speed depends on the direction of propagation of the wave or more accurately on the angle subtended between the wave vector and the spatial gradient of the field. In fact, when $\vec{k}$ is orthogonal to $\vec{\nabla} A$, the speed of gravity is equal to the speed of light. As we vary the angle the speed of gravity increases and it reaches its maximal possible value when $\vec{k}$ is aligned with $\vec{\nabla} A$.

Another important point is that both \eqref{sp1} and \eqref{sp2} allow for superluminal propagation. While the fundamental theory - General Relativity with a minimally-coupled scalar field, respects all the Lorenz symmetries, as the phase transition occurs and a non-minimal coupling forms, the Lorentz invariance is spontaneously broken. The reason behind that is that the BEC selects a preferred frame of reference and the gradient of the scalar field selects a preferred direction in spacetime. This Lorentz violation is not more drastic than what happens in theories of dark energy with non-minimal couplings and is also reminiscent of the Scharnhorst effect in optics \cite{scharn}. There higher order QED corrections modify the speed of light as it travels in the Casimir vacuum between two parallel tiny plates so that the speed in the direction orthogonal to the plates is greater than $c$. Because of the boundary condition set by the plates, the ground state breaks the Lorentz invariance and this leads to superluminal propagation even though the fundamental theory from which the effect is derived still respects all Lorentz symmetries. In fact, as shown in \cite{caus} the Scharnhorst effect does not lead to any causal paradoxes. This serves as an argument that also in the BEC model the superluminal propagation is not a problem and is consistent with all the assumptions that we have made.

It is an interesting question whether the modification of the propagation speed is the same for waves of all wavelengths. In fact, the answer is "no" and the easiest way to see that is through the dispersion relation $\omega^2 = c_{g}^2 k^2$. If we write the speed of gravity in the compact form
\begin{equation} 
c_{g}^2 = c^2 \Big( 1 + \frac{L^2 \alpha}{\delta^2} \Big)
\end{equation}
where $\delta$ is the length scale over which $\bar{\phi}$ changes and $\alpha$ is a dimensionless parameter which depends on the Planck length and the strength of the field, then the dispersion relation can be written as (keeping factors of $c$ just for clarity)
\begin{equation}
\omega^2 = c^2 k^2 + c^2 \Big( \frac{4 \pi^2 \alpha}{\delta^2} \Big) \frac{k^2}{\kappa^2}. 
\end{equation}
where in the second term on the RHS we have introduced $\kappa \equiv \frac{2 \pi}{L}$ in order to make the scaling between $k$ and $\kappa$ and hence $\lambda$ and $L$ more apparent. It is obvious that (keeping the gradient fixed) for $\kappa \gg k$, hence $\lambda \gg L$, this reduces to the standard relativistic dispersion relation, while a significant deviation from the standard relation occurs when $\lambda \sim L$ or $\lambda \ll L$. Thus the speed of propagation is modified only for wavelengths smaller than the healing length (which we have identified with the coupling constant). Since in our case the wavelength is much shorter than the healing length, as verified later, there will be a corresponding modification of the speed. On the other hand $\lambda$ has to be larger than the scattering length $a$ since on scales smaller than $a$ individual particles can be discerned and the condensate description breaks. Since the scalar field constituting dark matter is very weakly self-interacting, the scattering length has to be very small - much smaller than the wavelength of the observed gravitational waves. We elaborate more on this in Section \ref{other}.

An intriguing peculiarity of \eqref{sp1} and \eqref{sp2} is that they allow the denominator to become zero and therefore the speed to blow up. This is not a serious worry in our case because the terms proportional to the gradients of the scalar field can a posteriori be verified to be extremely small. Actually, when we later estimate the predicted difference in the arrival time of electromagnetic and gravitational waves, we will only need the linearly expanded versions of \eqref{sp1} and \eqref{sp2}, which can be compactly written as:
\begin{equation}
c_g^2 \approx 1 + 2 \Delta c_g
\end{equation}
where $\Delta c_g \equiv \frac{c_g - c}{c}$. For the cases of timelike and spacelike gradients, $\Delta c_g$ is given by
\begin{equation}
\Delta c_g \approx 8 \pi G L^2 | \dot{\bar{\phi}} |^2, \label{luther1}
\end{equation}
\begin{equation}
\Delta c_g \approx 8 \pi G L^2 | \bar{\phi}_{\parallel} |^2. \label{luther2}
\end{equation}
In the case when the scalar field is only radially dependent and its spatial gradient is aligned with the wave vector $\vec{k}$, the latter expression can be rewritten as:
\begin{equation}
\Delta c_g \approx 8 \pi G L^2 \vec{\nabla} A . \vec{\nabla} A = \frac{8 \pi G L^2}{m^2} \vec{\nabla} \sqrt{\rho} . \vec{\nabla} \sqrt{\rho} \label{lalala}
\end{equation}
Nevertheless, the potential divergence of \eqref{sp1} and \eqref{sp2} shows that these formulas cannot be applied at arbitrary large gradients of the scalar field. It is viable that there is a feedback mechanism built inside the model which prevents the building of such gradients.



\section{Constraining the BEC-DM Model with the GW-GRB Observation}
\label{time}


\subsection{Methodology and Assumptions}
\label{method}

We currently do not have a rigorous mathematical theory of condensation in a curved spacetime. As a result, there are several uncertainties in the model of BEC as DM. For example, we do not know how strong the gravitational potential wells need to be so that the condensate forms. Therefore, several scenarios are possible. The condensate might form only within galaxy halos, or it might also form outside galaxy halos. It might even be possible that the whole halo of a cluster of galaxies is in the condensed phase. We capture the uncertainty about the level at which the condensate forms by introducing a free parameter $\beta$ which is defined in the next subsection. 

A second source of uncertainty is whether the scalar field (i.e. the condensate wave function) oscillates or not around the minima of its potential. As we have argued before, in order for the scalar field to mimic CDM on large scales, the field has to oscillate. However, the presence of the non-minimal coupling $G_{\mu \nu} \nabla^{\mu} \phi^{*} \nabla^{\nu} \phi$ changes the evolution of the field with respect to the standard case. It is viable that this coupling would serve as a damping force which would tend to dissipate the oscillatory energy. In order to calculate the precise effect of the non-minimal coupling we would have to numerically solve for the evolution of the scalar field and also know details about the dynamics of the galaxy immediately after the condensation has happened and before is has reached equilibrium. Since this goes beyond the scope of the present work, we are going to capture the uncertainty about whether the oscillations are damped or not by introducing two extra free parameters  - $\gamma_{1}$ and $\gamma_{2}$. Including $m$ - the mass scale of the field, and $\varsigma \equiv \frac{1}{L}$ the scale of the non-minimal coupling, we have in total five free parameters - $m, \varsigma, \beta, \gamma_{1}, \gamma_{2}$. We will express the total predicted arrival time difference between GWs and GRBs that accumulates along the way between the source and the observer in terms of them.

In order to achieve that analytically, we work at an order of magnitude level. We also make several other simplifying assumption and approximations. We assume that the density distribution is given by the Burkert profile \eqref{Burkert} i.e. we assume that the quantum pressure has already smoothed out any cusps and the system has relaxed to its equilibrium profile (Reference \cite{bec} shows that one can obtain a profile similar to the Burkert one within this model). The dominant effect on the time delay of GRBs with respect to GWs comes from the halos of the Milky Way and the host galaxy and eventually from the halos of clusters between us and the event (all other halos are too far away from the physical path of the waves). We assume that the GWs and GRBs pass through the centre of each halo. At first, this might seem like a drastic assumption but it actually leads to the maximal possible difference in the arrival time of the two signals and is good enough for putting a constraint on the theory. Nonetheless, later we comment on the case where the waves pass at an impact parameter $b$. 

In order to perform an order of magnitude estimation of the difference in the arrival time, we also assume that in the case of spatial gradients the GW propagates with the maximal possible speed (i.e. the speed is constant and is calculated at the radius where the gradient of $\phi$ is largest) only inside the core of the halo (this agrees with a numerical calculation of the full arrival time difference performed for different values of the parameters $\varsigma$ and $m$), while in the case of time gradients, we later show that the modification of the speed of GWs is proportional to the average DM density of the halo. In both cases we work to first order in $\Delta c_g$ as given by Equations \eqref{luther1} and \eqref{luther2} respectively (a posteriori checks can verify that these terms are very small and numerical checks confirm that for the spatial case the first order calculation gives the same result as the full calculation).

We calculate the contribution to the arrival time difference coming from a single galaxy halo assuming that the galaxy is typical i.e. that it has the following parameters \cite{mo}:

\begin{center}
\begin{tabular}{ |c|c|c| } 
 \hline
 Dark matter density at the centre & $\rho_{0,g} = 3 \times 10^{-22} kg/m^3$ \\ 
 \hline
 Average dark matter density in the halo & $\bar{\rho}_{gh} = 10^{-23} kg/m^3$ \\ 
 \hline
 Radius of the core & $r_{0,g} = 15 kpc = 4.5 \times 10^{20} m$ \\ 
 \hline
 Radius of the halo & $R_{h,g} = 200 kpc = 6 \times 10^{21} m$ \\
 \hline
\end{tabular}
\end{center}

Similarly, when we take into account the possible contributions from clusters, we assume that they are typical with the following parameters:

\begin{center}
\begin{tabular}{ |c|c|c| } 
 \hline
 Dark matter density at the centre & $\rho_{0,cl} = 9 \times 10^{-23} kg/m^3$ \\ 
 \hline
 Average dark matter density in the halo & $\bar{\rho}_{cl} = 3 \times 10^{-24} kg/m^3$ \\ 
 \hline
 Virial radius & $R_{h,cl} = 1.2 \times 10^{23} m$ \\ 
 \hline
 Radius of the core & $r_{0,cl} = 2.4 \times 10^{22} m$ \\
 \hline
\end{tabular}
\end{center}

In order to calculate the average dark matter density for a galaxy halo and a cluster halo, we use the following formula which is obtained from numerical simulations \cite{lapi, huang}:
\begin{equation}
\bar{\rho} \approx 200 \rho_{cr}(z_{vir}),
\end{equation}
where $\rho_{cr}(z_{vir})$ is the critical density of the Universe evaluated at $z_{vir}$ and $z_{vir}$ is the redshift at which the galaxy or cluster virialized.

In order to obtain $r_{0,cl}$ from $R_{h,cl}$ we use the fact that they are related by \cite{locas}
\begin{equation}
r_{0,cl} = \frac{R_{h,cl}}{C_p},
\end{equation}
where the concentration parameter $C_p$ is approximately a constant (it is very weakly mass-dependent) and for a typical cluster of galaxies: $C_p \approx 5$.

Note that the above given $\rho_{0,cl}$ is obtained from $\rho_{0,g}$ by assuming they are related by
\begin{equation}
\frac{\rho_{0,cl}}{\rho_{0,g}} =  \frac{\bar{\rho}_{cl}}{\bar{\rho}_{gh}}.
\end{equation}
It is not certain whether this assumption holds - it is still debatable whether halos of clusters have a well-defined core and what their central density is. In any case, it is unlikely that this ambiguity will affect the order of magnitude constraints that we will later obtain.

The distance between the source and the observer is $\ell = 40 Mpc$. If we draw a line, between us and the source about half of the line would pass through a cluster. Since the typical virial radius of a cluster halo is about $4 Mpc$, then statistically there are about $5$ clusters or $N = 10$ half-clusters between us and the source.


\subsection{Parametrising the Model}

	The first source of uncertainty is whether the condensation occurs only within galaxy halos or also outside galaxy halos within the cluster halo. We started from the assumption that there is condensation within galaxies. What can we say about the physics at the scale of galaxy clusters? 
	
A priori, there is a range of different possibilities. It might be that there is no dark matter in the condensate phase outside galaxies, it might be that in some parts of the clusters where the gravitational fields are stronger the dark matter field has condensed or it might be that the whole cluster is in the condensate phase. In fact, we can capture all the different cases by introducing a term  $\tilde{L} G_{\mu \nu} \langle \phi \rangle ^ {\mu} \langle \phi \rangle ^{\nu}$ in the effective Lagrangian at cluster scales. Here $\langle \phi \rangle$ is the average of the field over cluster scales and $\tilde{L}$ is a new coupling constant which we keep as a free parameter that represents our uncertainty as to what exactly happens on cluster scales. 

We allow $\tilde{L}$ to have a range from $0$, which corresponds to the case where there is no condensation anywhere outside galaxies, to $\tilde{\xi} = \alpha \xi$ which corresponds to the case where the whole cluster is in the condensate phase. Here $\alpha$ is an empirical constant which takes into account the fact that a BEC in the whole cluster would have a different healing length than a BEC in a galaxy. In order to calculate $\alpha$ we need to take into account that the healing length of a condensate is related to the average matter density and radius of a halo by \eqref{healcore}. This implies
\begin{equation}
\alpha = \Big( \frac{\bar{\rho}_{cl}}{\bar{\rho}_{gh}} \Big)^{-1/2} \frac{R_{h,g}}{R_{h,cl}} \approx 0.1.
\end{equation}
We now define the parameter $\beta$ by
\begin{equation}
\beta \equiv \frac{\tilde{L}}{\alpha \xi}
\end{equation}
so that it has a range $(0, 1)$.

The second source of uncertainty is whether the field $\phi$, which corresponds to the condensate wave function, oscillates or not. This is important because any oscillatory behaviour would lead to large time gradients, which would contribute significantly to the modification of the speed of gravitational waves in a cosmological medium. The scalar field that constitutes Dark Matter would tend to oscillate. However, we saw that in our case the non-minimal coupling to the curvature could potentially serve as a damping force which would tend to relax this oscillation. The question is whether this damping really happens and how effective it is.

We can parametrise this uncertainty by introducing another free parameter $\gamma_{1}$. In the extreme case where there is no damping there would be simple harmonic oscillations with period proportional to the inverse of the mass of the scalar field. We model the effect of the damping by still assuming simple harmonic oscillations but with a longer period $T_{\mathrm{eff}}$ and therefore a smaller effective mass $m_{\mathrm{eff}}$. While this is not strictly true, it is good enough for the purposes of the present investigation. We define the dimensionless parameter $\gamma_{1}$ by
\begin{equation}
\gamma_{1} \equiv \frac{m_{\mathrm{eff}}}{m}.
\end{equation}
We see that it has limiting values $\gamma_{1} = 0$ corresponding to completely damped oscillations and $\gamma_{1} = 1$ corresponding to completely undamped (free) oscillations.

Next we need to take into account that if the condensation happens within clusters as well, then the oscillations might be damped to a different extent outside galaxy halos. In order to capture this uncertainty, we introduce another free parameter $\gamma_{2}$ which is defined in the same way as $\gamma_{1}$ but for oscillations within clusters, leaving $\gamma_{1}$ to represent only our uncertainty about oscillations within galaxies.


\subsection{Deriving the formula for the total arrival time difference}

The modification of the speed of propagation of gravitational waves with respect to the speed of light is proportional to the gradients of the scalar field (condensate wave-function). Larger gradients imply larger modification which in turn implies larger time lag between the two waves. Therefore the total time lag, the difference in arrival time which is detected at the observer, accumulates only when the waves pass through regions with large gradients of the field. There are four separate contributions to the arrival time difference that we need to consider - from spatial gradients within galaxies, from time gradients within galaxies, from spatial gradients within clusters, from time gradients within clusters. We will now derive formulas for each in turn.


\subsubsection{The contribution from spatial gradients in galaxies}

We first look at the contribution from the spatial gradients in galaxies. We make use of all the assumptions and approximations stated in Subsection \ref{method}.
The DM density distribution inside a galaxy halo is given by the Burkert density profile
\begin{equation}
\rho(r) = \frac{\rho_{0,g}}{(1 + \frac{r}{r_{0,g}})(1 + (\frac{r}{r_{0,g}})^2)}.
\end{equation}

Substituting that inside \eqref{lalala} one obtains:
\begin{equation}
\Delta c_g (r) = \frac{\rho_{0,g}}{4 M_p^2 \varsigma^2 m^2 r_{0,g}^2} F \Big( \frac{r}{r_{0,g}} \Big).
\end{equation}
where $M_p$ is the reduced Planck mass defined by
\begin{equation} 
M_{p} \equiv 1/\sqrt{8 \pi G} = 2.4 \times 10^{27} eV
\end{equation}
and $F$ is a function defined by
\begin{equation}
F(x) := \frac{(1 + 2x + 3x^2)^2}{(1+x)^3(1+x^2)^3}.
\end{equation}

The total time lag between electromagnetic and gravitational waves can be written as an integral over the line of sight
\begin{align}
\Delta t &= 2 \int^{\ell/2}_0 d r \Delta c_{g} (r) \nonumber \\
&= \frac{\rho_{0,g}}{2 M_p^2 \varsigma^2 m^2 r_{0,g}} \int^{\ell/2r_{0,g}}_0 F(x) d x. \label{timeint}
\end{align}

Since we assume that the wave propagates with the maximally modified speed over a distance equal to the diameter of the core, this reduces to
\begin{equation}
\Delta t_{g} = \frac{\rho_{0,g}}{4 M_p^2 \varsigma^2 m^2 r_{0,g}} F \Big( \frac{r_{\max,g}}{r_{0,g}} \Big). \label{tg}
\end{equation}
where $r_{\max,g}$ is the radius at which the gradient is maximal: $r_{\max,g} = 1.5 \times 10^{20} m$.


\subsubsection{The contribution from time gradients in galaxies}

Next we calculate the contribution from time gradients in galaxy halos. Since galaxies are virialized objects, they don't evolve very much. There are very small time gradients from rotation and fluxes of radiation and heat. However, they are mostly associated with the visible matter inside halos and their influence on the evolution of the dark matter field is negligible. The main contribution to the time gradient of the condensate field comes from possible oscillations of the field. Now we calculate this contribution in terms of the free parameter $\gamma_{1}$.
According to our assumptions, the field oscillates as a free field with effective mass $m_{\mathrm{eff}}$:
\begin{equation}
\bar{\phi} = A(r) \mathrm{e}^{i (m_{\mathrm{eff}} t + \alpha_{0})}.
\end{equation}
Therefore,
\begin{equation}
|\dot{\bar{\phi}}|^2 = m_{\mathrm{eff}}^2 A(r)^2 = m_{\mathrm{eff}}^2 \bar{\phi}^2 = \frac{m_{\mathrm{eff}}^2 \bar{\rho}_{gh}}{m^2} = \gamma_{1}^2 \bar{\rho}_{gh}.
\end{equation}
where $\bar{\phi}^2 \propto \bar{\rho}_{\mathrm{\mathrm{gh}}}$ is necessary since the field oscillates everywhere within the halo. The fractional modification of the speed of gravitational waves (written in terms of Newton's constant G) is
\begin{equation}
\Delta c_g = \frac{8 \pi G}{\varsigma^2} |\dot{\bar{\phi}}|^2 = \frac{8 \pi G \bar{\rho}_{\mathrm{gh}} \gamma_{1}^2}{\varsigma^2},
\end{equation}
where we make use of the assumed first order relation between the density and the scalar field. This immediately gives us the formula for the arrival time difference:
\begin{equation}
\Delta t_{g, osc} = 2 R_{h,g} \Delta c_g = \frac{16 \pi G R_{h,g} \bar{\rho}_{gh} \gamma_{1}^2}{\varsigma^2}.
\end{equation}


\subsubsection{The contribution from spatial gradients in clusters}

The dark matter density distribution inside a cluster halo follows the Burkert profile but with different values of the central density $\rho_{0,cl}$ and of the radius of the core $r_{0,cl}$: 
\begin{equation}
\rho(r) = \frac{\rho_{0,cl}}{(1 + \frac{r}{r_{0,cl}})(1 + (\frac{r}{r_{0,cl}})^2)}.
\end{equation}
As a consequence, the maximal value of the gradient of the scalar field is reached at $r_{max, cl} = 7.7 \times 10^{21} m$. The difference between the arrival times in this case is given by the same formula as \eqref{tg} but with the parameters of the cluster halo and with one major difference - the prefactor that we insert is, in accordance with our assumptions, $N \times r_{0,cl}$ where $N = 10$ is the number of half-clusters between us and the source:
\begin{equation}
\Delta t_{cl} = \frac{N \alpha^2 \beta^2 \rho_{0,cl}}{4 M_p^2 \varsigma^2 m^2 r_{0,cl}} F \Big( \frac{r_{max,cl}}{r_{0,cl}} \Big).
\end{equation}


\subsubsection{The contribution from time gradients in clusters}
The case of the time delay between the two waves arising from time gradients within clusters of galaxies mirrors almost exactly the case of time gradients within galaxies except that it is now proportional to the average dark matter density within clusters - $\bar{\rho}_{cl}$ and that the distance over which the GW propagates with modified speed is $\ell/2$:
\begin{equation}
\Delta t_{cl, osc} = \frac{4 \pi G \ell  \bar{\rho}_{cl} \alpha^2 \beta^2 \gamma_{2}^2}{\varsigma^2}.
\end{equation}


\subsubsection{The total arrival time difference}

Gathering all contributions together, we obtain a formula for the total predicted arrival time difference in terms of the five free parameters:
\begin{align}
\Delta t_{tot} &= \Delta t_{g} + \Delta t_{g,osc} + \Delta t_{cl} + \Delta t_{cl,osc} \nonumber \\
&= \frac{\rho_{0,g}}{4 M_p^2 \varsigma^2 m^2 r_{0,g}} F \Big( \frac{r_{\max,g}}{r_{0,g}} \Big) + \frac{16 \pi G R_{h,g} \bar{\rho}_{gh} \gamma_{1}^2}{\varsigma^2} \nonumber \\
&+ \frac{N \alpha^2 \beta^2 \rho_{0,cl}}{4 M_p^2 \varsigma^2 m^2 r_{0,cl}} F \Big( \frac{r_{max,cl}}{r_{0,cl}} \Big) + \frac{4 \pi G \ell \bar{\rho}_{cl} \alpha^2 \beta^2 \gamma_{2}^2}{\varsigma^2}. \label{ttot}
\end{align}
This is the main formula in our paper. It is easy to see how the total arrival time difference scales with the three parameters which parametrise our uncertainty - $\beta$, $\gamma_{1}$ and $\gamma_2$. It is proportional to the square of each parameter. Since each parameter has a range $(0, 1)$, the terms containing these parameters become important when the corresponding parameter gets close to $1$. This also means that we cannot a priori ignore any of these terms, since each of them becomes important in some regime. However, we can identify several physically important limiting cases and testing the model in each of these cases gives us an idea of the overall constraint.


\subsection{Testing the Limiting Cases}

There are two limiting values - 0 and 1, for each of the three parameters - $\beta$, $\gamma_{1}$ and $\gamma_2$, which naively would give a total of eight limiting cases.  However, two of them, $\{ \beta = 0, \gamma_{1} = 0, \gamma_{2} = 0 \}$ and $\{ \beta = 0, \gamma_{1} = 1, \gamma_{2} = 0 \}$ are redundant, since they correspond to no condensation outside galaxy halos, which in turn would make it irrelevant if the field there oscillates or not (in fact, it probably will, since there is no non-minimal coupling there to suppress the oscillations). 

Of the remaining six limiting cases, there are two classes of cases which lead to different results. It turns out that the constraint is not sensitive to the value of $\beta$, i.e. in the end it is not important whether the condensation happens only inside galaxy halos or also inside cluster halos. Instead the constraint is sensitive to the values of $\gamma_{1}$ and $\gamma_{2}$, i.e. it matters whether the oscillations are suppressed everywhere or they happen freely somewhere.

\paragraph{No field oscillations}

\begin{figure}[h!]
\centering
\includegraphics[width=0.5\textwidth]{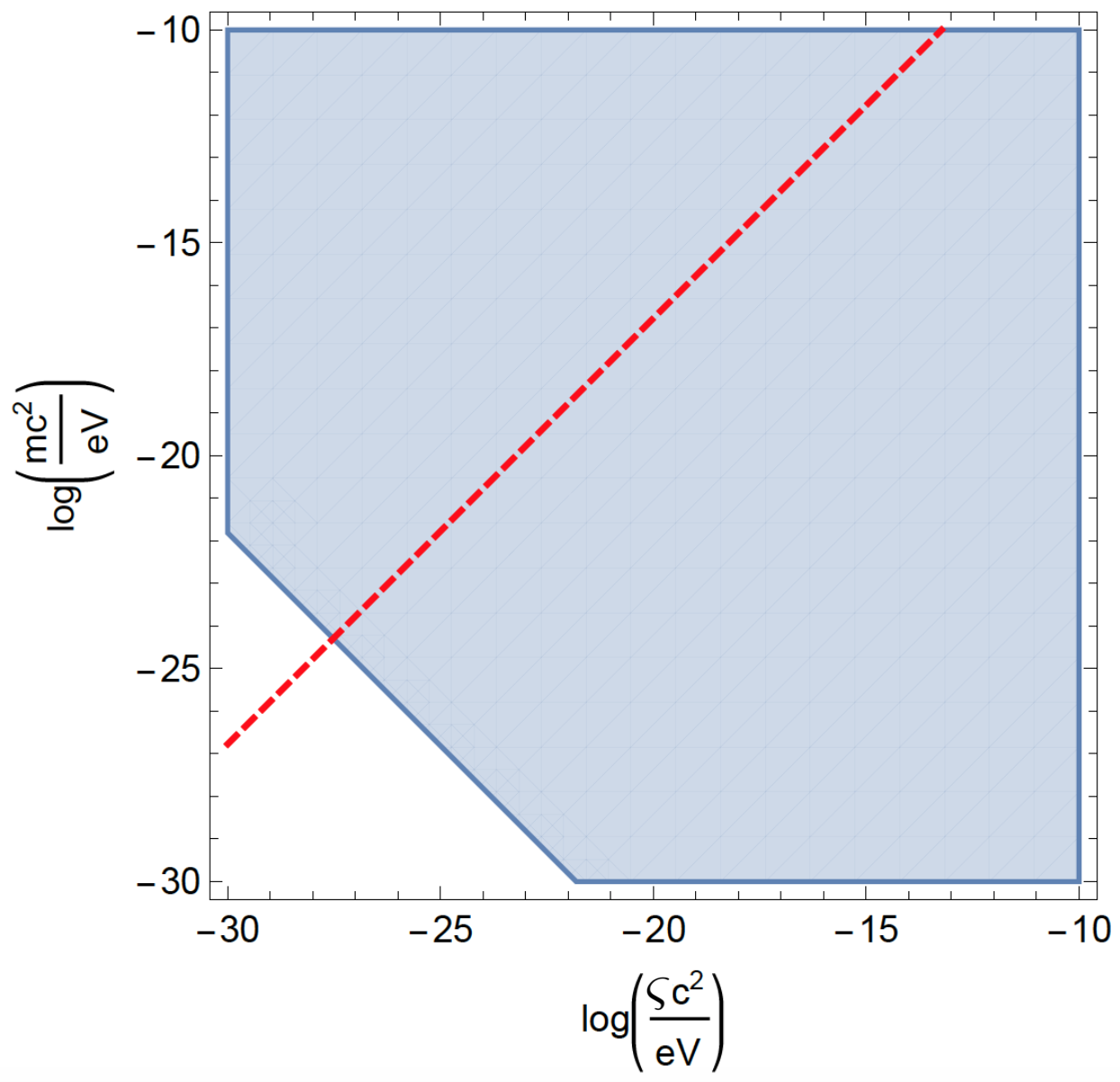}
\caption{\label{case1} A constraint on the $\varsigma-m$ parameter plane when the oscillations are completely suppressed everywhere. The arrival time difference between gravitational and electromagnetic waves is solely due to spatial gradients. Models which lie in the blue region are consistent with the observed time difference $\Delta t \lesssim 1.7 s$. The red line corresponds to values of $\varsigma$ and $m$ for which the BEC halo fits the size of a typical galaxy halo.}
\end{figure}

The first class of cases is when there are no oscillations anywhere in the condensate phase which corresponds to the cases $\{ \beta = 0, \gamma_{1} = 0 \}$ and $\{ \beta = 1, \gamma_{1} = 0,  \gamma_{2} = 0 \}$. Then the gradients are far too small to seriously affect the time difference between gravitational and electromagnetic waves. Calculating the constraint in $\varsigma - m$ space at an order of magnitude level, we obtain
\begin{equation}
\varsigma \times m \gtrsim 10^{-52} eV^2.
\end{equation}
Fig.\ref{case1} shows a constraint plot in the $\varsigma - m$ parameter space. The red line corresponds to the relation \eqref{mu} i.e. we identify the coupling constant $L$ with the healing length $\xi$ and demand that $\xi$ is of the order necessary to fit the size of non-relativistic BEC halo with the size of a typical DM halo. Thus it provides an independent constraint in the parameter space originating from the requirement that the density profile of the BEC fits with the observed density profile within galaxies. Combining this with the time of flight constraint, we get separate constraints on $\varsigma$ and $m$
\begin{equation}
m \gtrsim 10^{-24} eV, \qquad \varsigma \gtrsim 10^{-28} eV.
\end{equation}

\paragraph{Undamped field oscillations}

\begin{figure}[h!]
\centering
\includegraphics[width=0.5\textwidth]{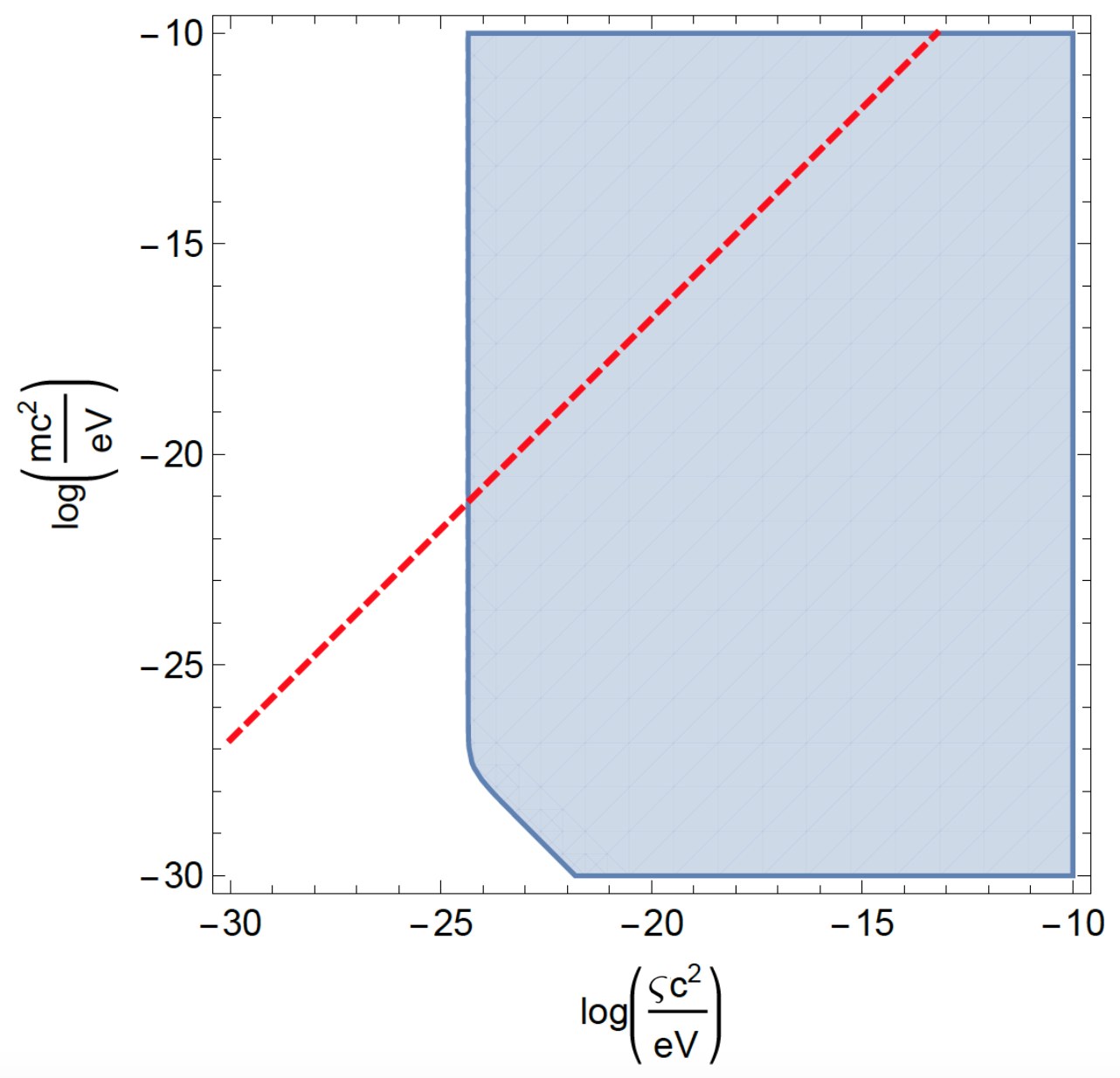}
\caption{\label{case2} A constraint on the $\varsigma-m$ parameter plane when the field oscillates somewhere in galaxy or cluster halos. The dominant contribution to the arrival time difference is from the time gradients. Models which lie in the blue region are consistent with the observed time difference $\Delta t \lesssim 1.7 s$.  The red line corresponds to values of $\varsigma$ and $m$ for which the BEC halo fits the size of a typical galaxy halo.}
\end{figure}

The second class of cases is when the oscillations happen somewhere - it could be either in galaxy halos, in cluster halos or in both. This corresponds to the cases $\{ \beta = 0, \gamma_{1} = 1 \}$, $\{ \beta = 1, \gamma_{1} = 1,  \gamma_{2} = 0 \}$, $\{ \beta = 1, \gamma_{1} = 0,  \gamma_{2} = 1 \}$ and $\{ \beta = 1, \gamma_{1} = 1,  \gamma_{2} = 1 \}$. Calculating the constraint in $\varsigma - m$ space at an order of magnitude level, we obtain a strong constraint on $\varsigma$
\begin{equation}
\varsigma \gtrsim 10^{-25} eV, \label{varsigma}
\end{equation}
and a weak constraint on $m$ coming from the spatial gradients of $\phi$ as calculated in the previous case. Fig. \ref{case2} shows a constraint plot in the $\varsigma - m$ parameter space. Again combining the time of flight constraint with the requirement that the BEC halo fits in size with a typical galaxy halo (the red line) leads to a separate constraint on $m$
\begin{equation}
m \gtrsim 10^{-21} eV.
\end{equation}
These constraints allow us to verify that $\lambda \ll L$ and therefore the GW is sensitive to the non-minimal coupling. Indeed, for $GW170817$, $\lambda \sim 10^5-10^7 m$, while \eqref{varsigma} implies that the region $L \gtrsim 10^{18} m$ is allowed. For small $L$ such that $L \sim \lambda$ the GW would no longer be sensitive to the non-minimal coupling and there would be no corresponding time delay between the two waves. However, since this region is allowed anyway, it does not change the constraint.

At first it might seem surprising that the contribution to the arrival time difference from cluster halos is of the same order as the contribution from galaxy halos especially in the case where the difference arises from the oscillations of the field. Naively, one would expect the contribution from cluster halos to be much larger due to the larger distance within a condensed region that the waves have to propagate (the factor of $\ell$ in the fourth term of \eqref{ttot} compared to the factor of $R_{h,g}$ in the second term). However, it turns out that this surplus in the propagation distance is exactly compensated by the fact that the healing length and hence the non-minimal coupling is of different size outside galaxy halos compared to inside (the factor of $\alpha^2$ in the fourth term). As a consequence, it is completely irrelevant for the time-of-flight constraint whether the condensation happens only inside galaxy halos or also outside. What matters only is whether the oscillations are suppressed or not. As we saw this uncertainty is captured by the parameter $\gamma_{1}$ (and $\gamma_{2}$ for oscillations outside galaxy halos). Formula \eqref{ttot} shows that the arrival time difference depends quadratically on $\gamma_{1}$ as it varies from $0$ to $1$. Thus as the oscillations are gradually switched on, larger and larger regions of the parameter space will get forbidden by the oscillations. Figures \ref{case1} and \ref{case2} show the limiting cases of $\gamma_{1} = 0$ and $\gamma_{1} = 1$.

\subsection{Waves at an impact parameter $b$}

So far we have only considered the case where the waves pass through the centre of the halo of a galaxy or a cluster. This is good enough for putting a constraint on the model because it maximises the total time difference accumulated along the path of the two waves. But how does the total arrival time difference change as we shift the path of the waves away from the centre of the halo while still preserving the spherical symmetry? This would only make a difference, if the arrival time difference is dominated by spatial gradients i.e. oscillations are largely suppressed. Then we can derive a formula for the arrival time difference in terms of the impact parameter $b$, which is defined as the closest distance between the path of the waves and the centre of the halo. Working again to first order in $\frac{1}{M_p^2 \varsigma^2} |\frac{d \bar{\phi}}{d r}|^2$, we obtain the formula which is just a slight modification of \eqref{timeint} above
\begin{equation}
\Delta t (b) = \frac{2}{M_p^2 \varsigma^2} \frac{l^2}{b^2 + l^2} \int^{l/2}_0 \left| \frac{d \bar{\phi}}{d r} \right|^2_{r = (b^2 + x^2)^{1/2}} d x.
\end{equation}
The difference between the arrival times of the two signals decreases quickly as $b$ is increased. For example, the nearest galaxy to the path of the waves, other than the Milky Way or the host galaxy, is at $b \sim 10^{22} m$ \cite{boran}. Taking values of $\varsigma$ and $m$ for which the model barely passes the test in the case of completely suppressed oscillations: $m \sim 10^{-24} eV$ and $\varsigma \sim 10^{-28} eV$ (i.e. the point where the red dashed line intersects the boundary between the allowed and forbidden regions in Fig.\ref{case1}) gives an additional time lag from the nearest galaxy of $\Delta t \sim 10^{-5} s$. This justifies our decision to ignore possible contributions to the arrival time difference from other galaxies.

\subsection{Other constraints on the mass}
\label{other}

The relation between the mass and the coupling constant \eqref{mu} which corresponds to the red dashed line in the plots suggests that we can increase both $m$ and $\varsigma$ indefinitely while still fitting the profile of a galaxy with that of a BEC halo. However, both $m$ and $\varsigma$ depend on the scattering length $a$ via \eqref{rad} and \eqref{heal} and the scattering length is not a completely free parameter in our model. Indeed, we can obtain an independent constraint on the mass $m$ by considering the possible range of $a$. 

The scattering length is a constant which characterises the strength of the interactions between the bosons in the condensate. Since the interactions have to be repulsive, $a$ has to be positive.\footnote{Incidentally, this implies that the boson which forms the condensate in this model cannot be the QCD axion for which the interactions are attractive \cite{qcd}. However, it could be a generalised axion particle.} The case $a = 0$ would correspond to no interactions. This is very unlikely since in order to form a condensate a weak interaction between the bosons is necessary (though see \cite{noint} for the possibility of forming a BEC without interactions). The case $a \lesssim l_{p}$ where $l_{p}$ is the Planck length would correspond to an interaction weaker than the gravitational one. This also seems unlikely since in that case gravitational interactions between the particles would dominate over their repulsive interactions. On the other hand, $a$ cannot become very large because this would imply strongly interacting dark matter which would be at odds with the observations. In particular since the lower limit for the range of observed wavelengths of the gravitational waves is $\lambda_{min} \sim 10^{5}$ $m$ and since a scattering length of that same order would imply very strong interactions of dark matter which is excluded by the observations, we can safely conclude that $a \ll \lambda$ and thus gravitational waves propagate safely in the regime where the condensate description holds. In fact, there is a hierarchy of length scales which goes like this:
\begin{equation}
l_{p} < a < \lambda < \xi .
\end{equation}
A possible, though highly overestimated, upper bound for $a$ is given by the scattering length of Rubidium atoms \cite{harko}: $a_{Rb} = 5.77 \times 10^{-9} m$. If we impose
\begin{equation}
l_{p} < a < a_{Rb},
\end{equation}
we obtain from Equations \eqref{rad} and \eqref{mu} a constraint on the mass $m$ given by
\begin{equation}
5.07 \times 10^{-10} \, \mathrm{eV} < m < 0.36 \, \mathrm{eV}.
\end{equation}
If we compare this interval to the time-of-flight constraints, we see that the interval lies deep into the allowed region of the parameter space both in the case of no oscillations and of free oscillations.



\section{Conclusion}
\label{conclusion}

In this paper we looked at the model of a relativistic non-minimally coupled BEC as a DM candidate and we tested it using the joint observation of $\mathrm{GW170817}$ and $\mathrm{GRB170817A}$. Since the model is mostly phenomenological, we had to add three free parameters $\beta$, $\gamma_{1}$ and $\gamma_{2}$ to capture our uncertainty of the scale at which the condensation occurs and of the behaviour of the non-minimally coupled field. In addition, we had two other free parameters - $\varsigma$ the scale of the non-minimal coupling and $m$ - the mass scale of the field. We derived an expression for the predicted arrival time difference in terms of these five parameters \eqref{ttot}, which is the main equation in our paper.  

We subsequently analysed the different limiting cases and found out that the constraint does not depend on whether the condensation happens only inside galaxy halos or also within cluster halos. However, it depends strongly on whether the scalar field (condensate wave function) oscillates or whether the oscillations are suppressed by the non-minimal coupling. 

In the case of completely damped oscillations we get only a very weak constraint in the $\varsigma - m$ parameter space (Fig.\ref{case1}), whereas in the case where unsuppressed oscillations happen somewhere on the path of the waves between the source and the observer, the constraint is much more stringent (Fig.\ref{case2}). Combining the time-of-flight constraint in both of these cases together with the requirement that the size of a theoretical BEC halo (which has no cusp because of the quantum pressure term) obtained in the minimally coupled non-relativistic limit fits with the size of a typical Dark Matter halo from observations, leads to a separate constraint on $\varsigma$ and m.

The constraints quoted above were obtained at an order of magnitude level. Further checks about how strong the model is constrained can be done only with numerical methods. We leave this for future investigations.


\section*{Acknowledgements}
We would like to especially thank Takeshi Kobayashi for constant support and feedback during the whole project. In addition, Dimitar Ivanov would like to thank Andrea Lapi, Dario Bettoni, Raul Carballo-Rubio, Paolo Salucci, Matteo Viel, Paolo Creminelli, Filippo Vernizzi and Diego Blas for useful discussions.


\appendix


\section{The Relation between $G_{\mu \nu} \nabla^{\mu} \phi \nabla^{\nu} \phi$ and $\mathcal{L}_4$ of Horndeski}
\label{equivalence}

Starting from $\mathcal{L}_5$ of the Horndeski Lagrangian with $G_5(\phi, X) = L^2 \phi$:
\begin{equation}
\mathcal{L}_5 = L^2 \phi G_{\mu \nu} \nabla^{\mu} \nabla^{\nu} \phi.
\end{equation}
Integrating by parts, ignoring the boundary term and using the Bianchi identity: $\nabla^{\mu} G_{\mu \nu} = 0$, leads to
\begin{equation}
\mathcal{L}_5 = - L^2 G_{\mu \nu} \nabla^{\mu} \phi \nabla^{\nu} \phi.
\end{equation}
Thus the coupling $L^2 G_{\mu \nu} \nabla^{\mu} \phi \nabla^{\nu} \phi$ is a subclass of the Horndeski action. Now we show that it is equivalent to $\mathcal{L}_4$ of Horndeski with $G_4(\phi, X) = L^2 X$:
\begin{equation}
\mathcal{L}_4 = L^2 XR - L^2 \big[ (\square \phi)^2 - \nabla_{\mu} \nabla_{\nu} \phi \nabla^{\mu} \nabla^{\nu} \phi \big]. \label{l4}
\end{equation}
Starting from $G_{\mu \nu} \nabla^{\mu} \phi \nabla^{\nu} \phi$ we have
\begin{align}
G_{\mu \nu} \nabla^{\mu} \phi \nabla^{\nu} \phi &= R_{\mu \nu} \nabla^{\mu} \phi \nabla^{\nu} \phi - XR \nonumber \\
&= \nabla_{\rho} \nabla_{\nu} \nabla^{\rho} \phi \nabla^{\nu} \phi - \nabla_{\nu} \nabla_{\rho} \nabla^{\rho} \phi \nabla^{\nu} \phi - XR \nonumber \\
& = - \nabla_{\nu} \nabla_{\rho} \phi \nabla^{\nu} \nabla^{\rho} \phi + (\square \phi)^2 - XR \nonumber \\
& = -\frac{1}{L^2} \mathcal{L}_4,
\end{align}
where in the first line we use the definition of the Einstein tensor, in the second line we use the definition of the Riemann tensor
\begin{equation}
R^{\rho}_{\mu \sigma \nu} V^{\mu} = \nabla_{\sigma} \nabla_{\nu} V^{\rho} - \nabla_{\nu} \nabla_{\sigma} V^{\rho}
\end{equation}
with $V^{\mu} = \nabla^{\mu} \phi$, in the third line we integrate by parts ignoring the boundary terms, and in the fourth line we use \eqref{l4}. Notice the minus sign that we have picked in the integration by parts. What we have essentially shown is that
\begin{equation}
\mathcal{L} = \frac{1}{16 \pi G} R + L^2 G_{\mu \nu} \nabla^{\mu} \phi \nabla^{\nu} \phi
\end{equation}
is equivalent to
\begin{equation}
\mathcal{L} = \frac{1}{16 \pi G} R - L^2 \Big\{ XR - \big[ (\square \phi)^2 - \nabla_{\mu} \nabla_{\nu} \phi \nabla^{\mu} \nabla^{\nu} \phi \big] \Big\}.
\end{equation}




\end{document}